\documentclass[prl,twocolumn,floatfix,nobibnotes,superscriptaddress,prb]{revtex4}
\usepackage{amsmath}
\usepackage{graphicx}

\begin{document}
\title{Phonon Life-times from first principles self consistent lattice dynamics}
\author{Petros Souvatzis}
\affiliation{Department of Physics and Astronomy, Division of Materials Theory Uppsala University,
Box 516, SE-751210, Uppsala, Sweden}

\begin{abstract}
Phonon lifetime calculations from first principles usually rely on time consuming molecular dynamics calculations,
or density functional perturbation theory (DFPT) where the zero temperature crystal structure is assumed to be dynamically stable.
Here is presented a new and effective method for calculating phonon lifetimes from first principles. This method is not limited to crystallographic phases  stable at 0 K, and provides a scheme 
more effective than most corresponding molecular dynamics calculations. The method is based on the recently developed self consistent {\it ab initio} lattice dynamical method (SCAILD) and is here tested by calculating 
the bcc phase phonon lifetimes of Li, Na, Ti and Zr, as representative examples.
\end{abstract}

\maketitle
Calculations of phonon lifetimes are of great interest to the materials science community, partly in that 
it allows for the prediction of  thermal transport properties, which in themselves are important from an engineering perspective, but also,  
since these properties are closely interrelated to perhaps the even more intriguing thermoelectric effect \cite{chunlei}.

Historically phonon lifetime calculations from first principles have been performed for quite some time, for instance see the work of Katsnelson {\it et al.} \cite{kats} or Debernardi {\it et al.} \cite{Debernardi}.
More recently Koker \cite{koker} successfully  calculated the thermal transport coefficients for MgO 
within the context of {\it ab initio} molecular dynamics (ab-MD) \cite{PARINELLO}, and Bonini {\it et al.}  estimated  the optical phonon line-widths in graphene using density functional perturbation theory 
(DFPT) \cite{DFPT1,DFPT2,DFPT3}.

 Up to present date  ab-MD and DFPT  have been the  two main methods of choice used for calculating phonon lifetimes from first principles.
However, the use of ab-MD suffers from the long simulation times  required to sample the
correlation functions, whereas DFPT is limited to crystal structures stable at zero temperature, in that it only 
calculates third order anharmonic contributions. Thus, there is still a need for more effective methods, preferably  based upon 
schemes that  also  allow lifetime calculations in crystallographic phases stabilized only at finite temperatures, such 
as the bcc phase of Ti,  Zr and Hf and  the cubic phases of numerous shape memory alloys. 

In this paper a novel and effective method for calculating phonon-lifetimes, which do not suffer  from the limitations of ab-MD
and DFPT, will be presented. The method is based on the recently developed self consistent ab initio lattice dynamical scheme 
(SCALD)\cite{petros1}, which has previously successfully  been used in predicting phonon frequency renormalization  by strong anharmonicity \cite{petros11,petros12}.
The newly developed method will here be used to calculate phonon lifetimes in the bcc phase of Li, Na, Ti and Zr, in order to illustrate its effectiveness.

The SCAILD method is based on the calculation of Hellman-Feynman
forces of atoms in a supercell. The  method can be viewed as an
extension of the frozen phonon method \cite{FP1}, in which all
phonons with wave vectors $\mathbf{q}$ commensurate with the
supercell are excited together in the same cell by displacing
atoms situated  at the undistorted positions
$\mathbf{R}+\mathbf{b}_{\sigma}$, according to
 $\mathbf{R}+\mathbf{b}_{\sigma}  \rightarrow \mathbf{R}+\mathbf{b}_{\sigma} + \mathbf{U}_{\mathbf{R}\sigma}$, where the displacements are given by
\begin{equation}
\mathbf{U}_{\mathbf{R}\sigma}
= \frac{1}{\sqrt{N}}\sum_{\mathbf{q},s}\mathcal{A}_{\mathbf{q}s}^{\sigma}
\mathbf{\epsilon}_{\mathbf{q}s}^{\sigma}e^{i\mathbf{q}(\mathbf{R}+\mathbf{b}_{\sigma})}.\label{eq:SUPERPOS}
\end{equation}
Here $\mathbf{R}$ represent the $N$ Bravais lattice sites of the
supercell, $\mathbf{b}_{\sigma}$ the position of atom $\sigma$
relative to the lattice site, $\mathbf{\epsilon}_{\mathbf{q}s}^{\sigma}$
are the phonon eigenvectors corresponding to the phonon mode, $s$,
and the mode amplitude $\mathcal{A}_{\mathbf{q}s}^{\sigma}$  is calculated from the
different phonon frequencies
 $\omega_{\mathbf{q}s}$ through
\begin{equation}
 \mathcal{A}_{\mathbf{q}s}^{\sigma} =\pm \sqrt{
\frac{\hbar}{M_{\sigma}\omega_{\mathbf{q}s}} \Big
(\frac{1}{2}+n_{\mathbf{q}s}\Big )},
\label{eq:AMPL}
\end{equation}
where $n_{\mathbf{q}s}=n(\frac{\omega_{\mathbf{q}s}}{k_{B}T})$, with $n(x)=1/(e^{x}-1)$, are the phonon 
occupational numbers, $M_{\sigma}$ the atomic masses and $T$ is the temperature of the system. The phonon frequencies, $\omega_{\mathbf{q}s}$, are obtained  through 
the projections of the Fourier transformed atomic forces, $\mathbf{F}_{\mathbf{q}}^{\sigma}$, onto the eigenvectors of the corresponding mode
\begin{eqnarray}\label{eq:FOURIER}
\omega _{\mathbf{q}s}^{2} =\sum_{\sigma}\frac{\mathbf{\epsilon}_{\mathbf{q}s}^{\sigma} \cdot \mathbf{F}_{\mathbf{q}}^{\sigma}}{\mathcal{A}_{\mathbf{q}s}^{\sigma}M_{\sigma}},
\end{eqnarray}

Due to the simultaneous presence of all the commensurate phonons
in the same force calculation, the interaction between different
lattice vibrations are  taken into account and the phonon
frequencies given by  Eq. (\ref{eq:FOURIER}) are thus renormalized
by the very same interaction.

By alternating between calculating the forces on the displaced
atoms and calculating new phonon frequencies and new displacements
through Eqs. (\ref{eq:SUPERPOS})-(\ref{eq:FOURIER}) the phonon
frequencies are calculated in a self consistent manner. 

During the course of a SCAILD calculation frequency distributions, $\mathcal{D}(\omega_{\mathbf{q},s}^{2})$, are generated 
for each of the modes having q-vectors commensurate with the supercell \cite{petros2,petros3}. 
By calculating the first  moment, $\Omega _{\mathbf{q}s}^{2}$, of these distributions  the renormalized phonon frequencies are obtained \cite{petros1}.

The novel implementation of the SCAILD scheme utilizes the possibility of extracting the phonon lifetimes from the 
frequency distributions, $\mathcal{D}(\omega_{\mathbf{q},s}^{2})$. To establish the connection between the phonon lifetime, $\tau_{\mathbf{q}s}$, of 
a mode and its corresponding distribution $\mathcal{D}(\omega_{\mathbf{q},s}^{2})$, we start from the newtonian equations of motion  
 of a stochastically damped phonon mode
\begin{equation}
\frac{d^{2}\mathcal{A}_{\mathbf{q}s}^{\sigma}}{dt^{2}}= -\Gamma_{\mathbf{q}s}\frac{d\mathcal{A}_{\mathbf{q}s}^{\sigma}}{dt}-\Omega_{\mathbf{q}s}^{2}\mathcal{A}_{\mathbf{q}s}^{\sigma}+a(t),
\label{eq:stoch}
\end{equation}
where $\Gamma_{\mathbf{q}s}=1/\tau_{\mathbf{q}s}$ is the phonon linewidth (damping), and $a(t)$ is the stochastic acceleration arising from the interaction 
between the mode and the other phonons of the system. The replacement of  interaction terms by stochastic variables, is in  path-integral approaches associated with the  Stratonovich-Hubbard transformation \cite{strat}, and 
in more classical contexts, associated with discussions concerning the fluctuation dissipation theorem \cite{kubo}.

Following a path similar to what is  generally used in proving the  fluctuation dissipation theorem \cite{kubo,mcquarrie}, the damping $\Gamma_{\mathbf{q}s}$
is formally  connected to the acceleration correlation function, $C(t)\equiv\langle a(t_{0})a(t_{0}+t)\rangle$, through the integral equation
\begin{equation}
\Gamma_{\mathbf{q}s} = \frac{2M_{\sigma}}{3K_{B}T}\Big (\frac{\Omega_{\mathbf{q}s}}{2M_{\sigma}\nu_{\mathbf{q}s}}\Big )^{2}C^{*}(\Gamma_{\mathbf{q}s}),
\label{eq:gammaa}
\end{equation}
where the brackets $\langle,\rangle$ denote an ensemble average, $\nu_{\mathbf{q}s}\equiv \Omega_{\mathbf{q}s}^{2}-(\Gamma_{\mathbf{q}s}/2)^2$ and $C^{*}$ is given by
\begin{eqnarray}
C^{*}(\Gamma_{\mathbf{q}s}) =\nonumber \qquad \qquad \qquad \qquad \\
 \int_{-\infty}^{\infty}dt \Big [cos(\nu_{\mathbf{q}s}t)-\alpha_ {\mathbf{q}s}cos(\nu_{\mathbf{q}s}|t|+\delta_{\mathbf{q}s})\Big]C(t)e^{-\Gamma_{\mathbf{q}s}|t|}.
\label{eq:gamma}
\end{eqnarray}
Here $tan(\delta_{\mathbf{q}s})=2\nu_{\mathbf{q}s}/\Gamma_{\mathbf{q}s}$ and $\alpha_{\mathbf{q}s} = (\Gamma_{\mathbf{q}s}/2)\Omega_{\mathbf{q}s}^{-1}$.

Even though Eq. (\ref{eq:gammaa}-\ref{eq:gamma}) in principle  provides 
a connection between the stochastic friction forces and the phonon lifetimes, we will here instead follow an alternative approach in order to extract the phonon lifetimes,
which is more convenient when employed in the context of the  SCAILD scheme. 

The phonon frequencies obtained through the SCAILD scheme, Eq. (\ref{eq:FOURIER}), can be related to the homogeneous (transient)  solutions to  
Eq. (\ref{eq:stoch}), i.e solutions achieved with $a(t)=0$, through
\begin{equation}
\omega _{\mathbf{q}s}^{2} = -\frac{1}{\mathcal{A}_{\mathbf{q}s}^{\sigma}}\frac{d^{2}\mathcal{A}_{\mathbf{q}s}^{\sigma}}{dt^{2}} = 
\frac{\Gamma_{\mathbf{q}s}}{\mathcal{A}_{\mathbf{q}s}^{\sigma}}\frac{d\mathcal{A}_{\mathbf{q}s}^{\sigma}}{dt}+\Omega_{\mathbf{q}s}^{2}.
\label{eq:gamma2}
\end{equation}
By assuming that the temperature is high enough  for a classical description to be valid, we can use the transient  solutions of Eq. (\ref{eq:stoch}) together 
with Eq. (\ref{eq:gamma2})  to obtain the following relation
\begin{equation}
\mathcal{D}(\omega_{\mathbf{q},s}^{2}) \sim \Big | \frac{d \omega _{\mathbf{q}s}^{2}}{dt} \Big |^{-1}=\frac{\Gamma_{\mathbf{q}s}}{ (\Gamma_{\mathbf{q}s}
\omega_{\mathbf{q},s})^{2}+(\omega_{\mathbf{q},s}^{2}-\Omega_{\mathbf{q}s}^{2})^{2}}
\label{eq:distr}
\end{equation}
between the frequency distributions and the phonon lifetimes. 
Thus it  becomes evident  that the distribution (\ref{eq:distr}) calculated within the  SCAILD scheme corresponds, within a multiplicative constant, to the dynamical structure 
factor of a stochastically damped harmonic oscillator in the high temperature classical limit \cite{harmsto1,harmsto2}. 

Another important observation to make at this point,
is that the frequency distributions can be calculated despite the absence of any explicit time dependence in the SCAILD scheme, 
by virtue of the stochastic damping  and the ergodic principle.

Regarding the other computational details of the force
calculation the VASP package \cite{VASP} was used, within the
generalized gradient approximation (GGA). The PAW potentials
required energy cutoffs of 210 eV, 125 eV, 200 eV, 170 eV for Li, Na, Ti and Zr, respectively. The k-point mesh used
was $6\times6\times6$ Monkhorst-Pack, together with a Methfessel Paxton smearing of 0.2 eV. The 
supercells used were obtained by increasing the bcc primitive cells
8 times along the primitive lattice vectors, resulting in 512 atom cells. 

The phonon frequency distributions were obtained 
through 200 SCAILD iteration  together with a gaussian smearing of 0.05 THz$^{2}$ applied to each of the sampled 
squared frequencies.

\begin{figure}[tbp]
\begin{center}
\includegraphics*[angle=0,scale=0.35]{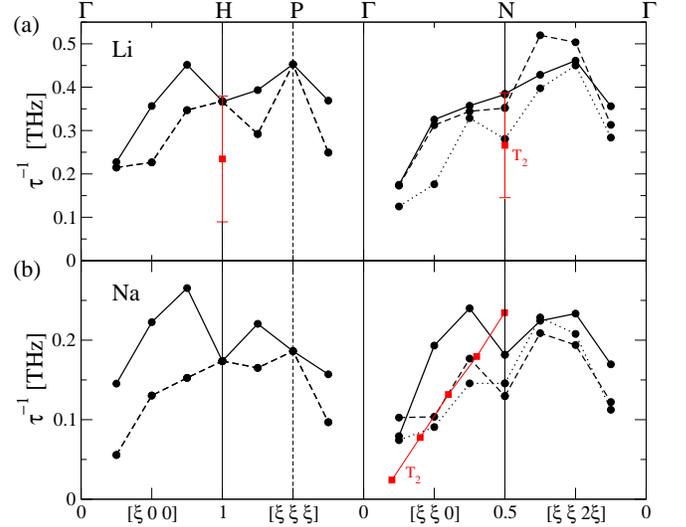}
\caption{(Color online) Calculated phonon lifetimes (full black circles) for (a) bcc-Li and (b) bcc-Na,  
displayed together with experimental data (red squares)\cite{ExpLi,ExpNa1,ExpNa2}. The full lines connect  
lifetimes corresponding to longitudinal modes, the dashed lines connect lifetimes corresponding to the
transverse T$_{1}$ modes and the dotted line connects lifetimes corresponding to the transverse T$_{2}$ modes.
The calculations were performed at  a temperature of 293 K , whereas the experimental data for bcc-Li and bcc-Na
were obtained at 293 K and 296 K, respectively.}
\label{fig:LiNa1}
\end{center}
\end{figure}
\begin{figure}[tbp]
\begin{center}
\includegraphics*[angle=0,scale=0.63]{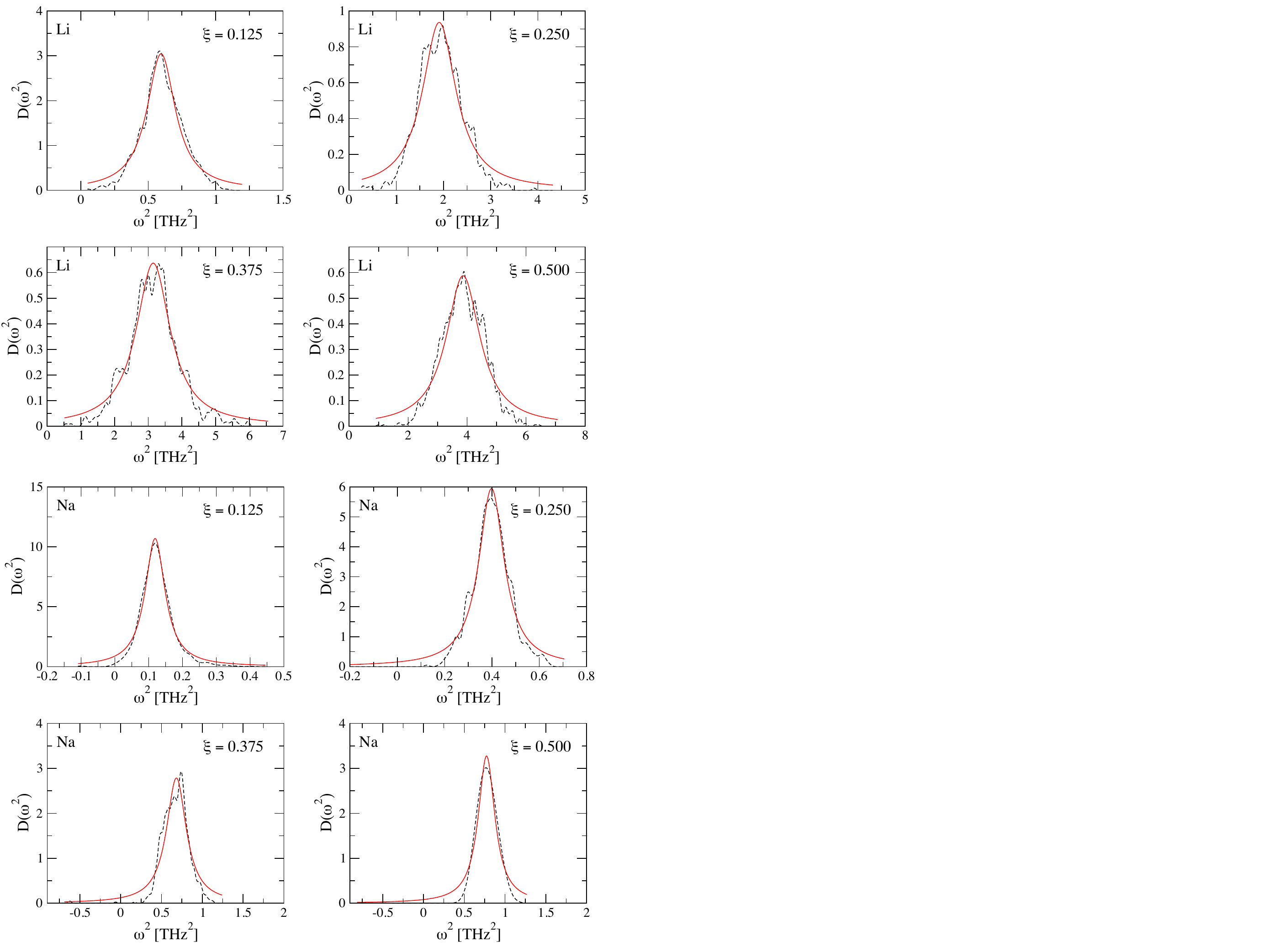}
\caption{(Color online) The calculated T$_{1}$ mode frequency distribution for  bcc-Li and bcc-Zr
along the $[\xi \xi 0]$ direction, for $\xi=0.125,0.250,0.375,0.500$ (dashed black lines). The full red curves were obtained through the 
fitting of the data with Eq. (\ref{eq:distr}). The calculations were performed at the finite temperature of 293 K. }
\label{fig:DENSLiNa}
\end{center}
\end{figure}
\begin{figure}[tbp]
\begin{center}
\includegraphics*[angle=0,scale=0.35]{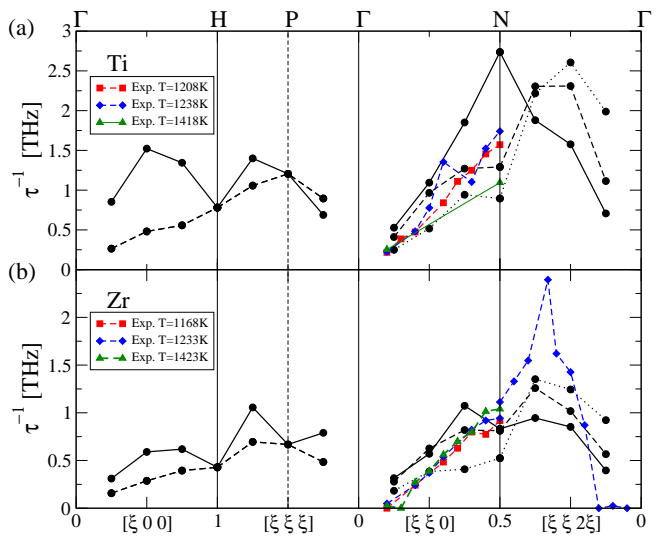}
\caption{(Color online) Calculated phonon lifetimes (full black circles) for (a) bcc-Ti and (b) bcc-Zr, 
displayed together with experimental data (red squares, blue diamonds and green triangles)\cite{ExpTi,ExpZr}. The full lines connect  
lifetimes corresponding to longitudinal modes, the dashed lines connect lifetimes corresponding to the
transverse T$_{1}$ modes and the dotted line connects lifetimes corresponding to the transverse T$_{2}$ modes.
The calculations were performed at  a temperature of 1300 K, whereas the temperature for which the experimental data were 
obtained is given in the figure. All the experimental lifetimes correspond to transverse T$_{1}$ modes.}
\label{fig:TiNa}
\end{center}
\end{figure}
\begin{figure}[tbp]
\begin{center}
\includegraphics*[angle=0,scale=0.63]{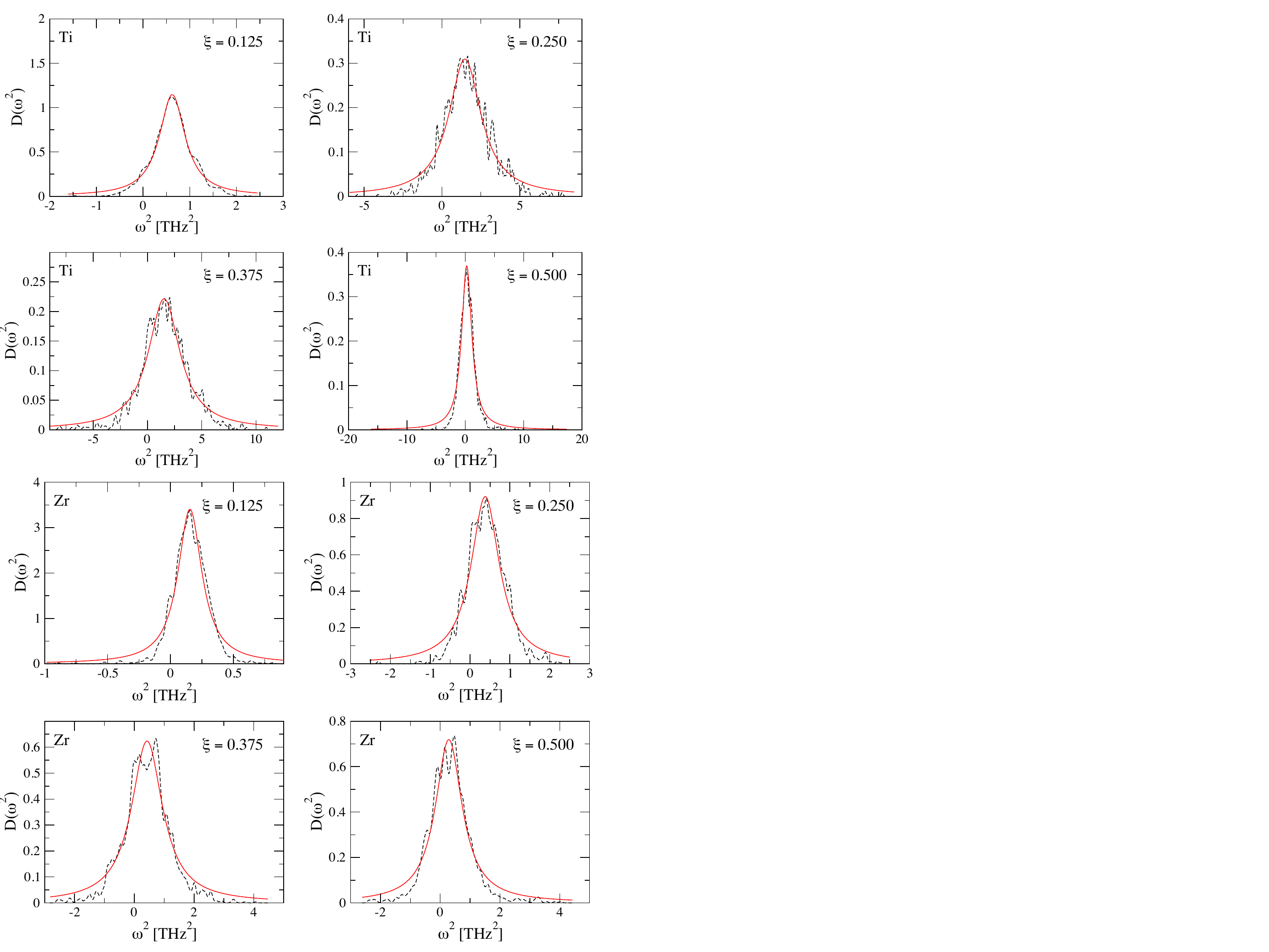}
\caption{(Color online) The calculated T$_{1}$ mode frequency distribution for  bcc-Ti and bcc-Zr
along the $[\xi \xi 0]$ direction, for $\xi=0.125,0.250,0.375,0.500$ (dashed black lines). The full red curves were obtained through the 
fitting of the data with Eq. (\ref{eq:distr}). The calculations were performed at the finite temperature of 1300 K. }
\label{fig:DENSTiZr}
\end{center}
\end{figure}

In Fig. \ref{fig:DENSLiNa} and Fig. \ref{fig:DENSTiZr} the calculated frequency distributions of bcc-Li and bcc-Ti are presented  together 
with their respective fits to Eq. (\ref{eq:distr}). Here  the distributions obtained from the SCAILD iterations are found to be in excellent agreement with the fit to  Eq. (\ref{eq:distr}).
Furthermore, in Fig. \ref{fig:LiNa1} and Fig. \ref{fig:TiNa} the phonon lifetimes extracted from the fits to Eq. (\ref{eq:distr}) are displayed together with experimental 
data. Here good agreement between theory  and experiment can be found in the $[110]$ direction for bcc-Na, bcc-Ti and bcc-Zr, whereas a potentially  quite huge discrepancy
can be found for bcc-Zr in the $[112]$ direction at $\xi=0.35$. This discrepancy is referred to as potential due to the fact that the experimental q-point corresponding to
$\xi=0.35$ does not belong to the q-point set used in the calculation, and because linear interpolation between lifetimes corresponding to different q-points is not always 
a good approximation to employ.   Finally, it can also be observed that  the theoretical phonon lifetimes in the case of bcc-Li are well within the experimental error bars.

In summary, a novel, effective and simple method for calculating phonon lifetimes have been presented together with some test calculations on the bcc phase occurring at 
room temperature as well as for temperatures well above 1000 K,  for Li , Na,  Ti and Zr. The method has provided good estimates to most of the 
phonon lifetimes, and show promise as an effective tool in calculating phonon line-widths, especially  for  crystal structures that 
can only be stabilized at elevated temperatures. Thus opening the door to phonon lifetime calculations from first principles  in cases when  anharmonic interactions 
of order $>$ 3 are important for achieving a correct description of the system under study.
\newline

I would like to thank the Swedish National Infrastructure for
Computing (SNIC) for making this work possible through  the allocation of computational time to account SNIC 006/10-3 at the PDC Lindgren cluster.

\end{document}